\documentclass[a4paper]{jpconf}
\usepackage{epsfig,bm,amssymb}
\begin{document}
\title{Local control of Hamiltonian chaos}

\author{C Chandre, G Ciraolo, R Lima, M Vittot}

\address{Centre de Physique Th\'eorique\footnote{Unit\'e Mixte de Recherche (UMR 6207) du CNRS, et des universit\'es Aix-Marseille I, Aix-Marseille II et du Sud Toulon-Var. Laboratoire affili\'e \`a la FRUMAM (FR 2291).}, CNRS Luminy, Case 907, F-13288 Marseille Cedex 9,
France}

\ead{chandre@cpt.univ-mrs.fr}

\begin{abstract}
We review a method of control for Hamiltonian systems which is able to create smooth invariant tori. This method of control is based on an apt modification of the perturbation which is {\em small} and {\em localized} in phase space. 
\end{abstract}

\section{Introduction}

\indent Controlling chaotic transport is a key challenge in many branches of physics like for instance, in particle accelerators, free electron lasers or in magnetically confined fusion plasmas.
For these systems, it is essential to control transport properties without significantly
altering the original system under investigation nor its overall chaotic structure. Here we review a control strategy for Hamiltonian systems which is based on building barriers by adding a small apt perturbation which is localized in phase space, hence confining all the trajectories. For more details on the methods (including exact results and numerical implementations, we refer to Refs.~\cite{guido1,guido2,michel,guido3,vitt04}).

We consider the class of Hamiltonian systems that can be written
in the form $H=H_0+V$ i.e.\ an integrable 
Hamiltonian $H_0$ (with action-angle variables)
plus a small perturbation $V$. Generically, it is  expected that the phase space of $H$ is a mixture between regular and chaotic behaviors. The transition to Hamiltonian chaos occurs by successive break-ups of invariant KAM tori which foliate the phase space of $H_0$. For two degrees of freedom, it is worthwhile noticing that this transition is very similar to phase transitions in statistical mechanics, in the sense that it can be described by hyperbolic invariant sets of a renormalization operator~\cite{Bmack93,esca85,PRchan}.

More than 25 years ago, Chirikov~\cite{chir79} stated an empirical criterion based on the overlap of primary resonances in order to get an idea on how chaos arises in Hamiltonian systems and to get rough estimates of the threshold of large scale diffusion. The idea of Chirikov is that chaos is obtained when the primary resonant islands overlap, i.e., when the distance between two of them is smaller than the sum to their half-widths. We now consider a small modification of the original system $H_c=H_0+V+f$ (where $f$ is smaller than $V$ with an appropriate norm of functions). The perturbation $f$ introduces generically additional resonances. Therefore from Chirikov's criterion, it is expected that there will be more overlaps of resonant islands and hence more chaos. 

The aim of our control strategy is to tailor appropriate perturbations $f$ acting in the opposite direction, namely such that the Hamiltonian $H_c=H_0+V+f$ has more smooth invariant tori and hence is more regular than $H=H_0+V$ contrary to the conventional wisdom inherited from Chirikov's criterion. For practical purposes, the control term should be small
with respect to the perturbation $V$, and localized in phase space
(i.e.\ the subset of phase space where $f$ is non-zero is finite
and small).

In this article, we highlight a method of control on a specific and paradigmatic example, the forced pendulum. We compute explicitly the formula of the control term which is able to create an isolated barrier of transport. We show that not only our method of control provides a control term which is much smaller than the perturbation but it also provides the explicit location of the created invariant torus which is a crucial step in order to localize the action of the control.

\section{Statement of the localized control theory}

\begin{figure}
\begin{center}
\epsfig{file=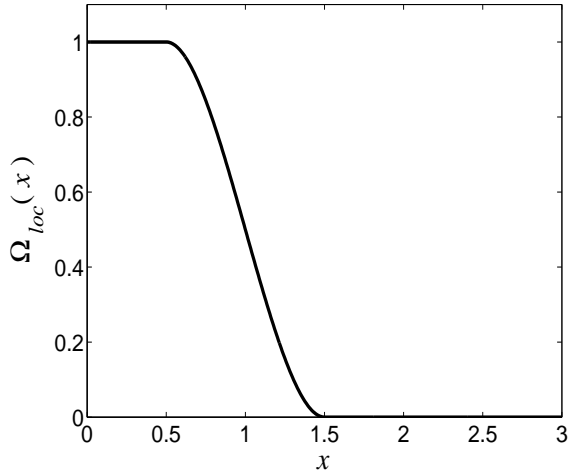,width=7.5cm,height=6.3cm}
\end{center}
\caption{$\Omega_{loc}(x)$ for $\alpha=0.5$ and $\beta=1.5$.}
\label{fig:omegaloc}
\end{figure}

The localized control method has been extensively described in Ref.~\cite{vitt04} where the corresponding rigorous mathematical results were proved. We state here the main result of this paper.
For a Hamiltonian system written in action-angle variables with $L$ degrees of freedom, the perturbed Hamiltonian is 
$$
H({\bf A},{\bm\theta})={\bm \omega}\cdot {\bf A}+ V({\bf A},{\bm\theta}),$$ 
where $({\bf A},{\bm\theta})\in {\mathbb R}^L\times {\mathbb T}^L$ and $\bm\omega$ is a non-resonant vector of ${\mathbb R}^L$. 
Without loss of generality, we consider a region near ${\bf A}={\bf 0}$ (by translation of the actions) and, since the Hamiltonian is nearly integrable, the perturbation $V$ has constant and linear parts in actions of order $\varepsilon$, i.e.\ 
\begin{equation}
\label{eqn:e4V}
V({\bf A},{\bm\theta})=\varepsilon v({\bm\theta})+\varepsilon {\bf w}({\bm\theta})\cdot {\bf A}+Q({\bf A},{\bm\theta}),
\end{equation} 
where $Q$ is of order $O(\Vert {\bf A}\Vert ^2)$. We notice that for $\varepsilon=0$, the Hamiltonian $H$ has an invariant torus with frequency vector ${\bm\omega}$ at ${\bf A}={\bf 0}$ for any $Q$ not necessarily small.
The controlled Hamiltonian we construct is
\begin{equation}
\label{eqn:gene}
H_c({\bf A},{\bm\theta})={\bm \omega}\cdot {\bf A}+ V({\bf A},{\bm\theta})+  f({\bm \theta})\Omega( {\bf A}, {\bm\theta} ),
\end{equation}
where $\Omega$ is a smooth characteristic function of a region around a targeted invariant torus. We notice that the control term $f$ we construct only depends on the angle variables and is given by
\begin{equation}
\label{eqn:exf}
f({\bm\theta})=V({\bf 0},{\bm\theta})-V\left( -\Gamma \partial_{\bm\theta} V({\bf 0},{\bm\theta}),{\bm\theta}\right),
\end{equation}
where $\Gamma$ is a linear operator defined as a pseudo-inverse of ${\bm\omega}\cdot \partial_{\bm\theta}$, i.e.\ acting on $V=\sum_{{\bf k}}V_{{\bf k}} {\mathrm e}^{i{\bm\theta}\cdot{\bf k}}$ as
$$
\Gamma V=\sum_{{\bm\omega}\cdot{\bf k}\not= 0} \frac{V_{{\bf k}}}{i{\bm\omega}\cdot{\bf k}} {\mathrm e}^{i{\bm\theta}\cdot{\bf k}}.
$$
Note that $f$ is of order $\varepsilon^2$. This can be seen from Eq.~(\ref{eqn:e4V}) since $f$ can be rewritten as
$$
f({\bm\theta})=\varepsilon^2 {\bf w}({\bm\theta})\cdot \Gamma \partial_{\bm\theta} v-Q\left(  -\varepsilon \Gamma \partial_{\bm\theta} v, {\bm\theta} \right),
$$
and $Q$ is quadratic in the actions. Here the dependence of the control in the actions ${\bf A}$ is in the function $\Omega$.
For any perturbation $V$, Hamiltonian~(\ref{eqn:gene}) has an invariant torus with frequency vector close to ${\bm\omega}$.
The equation of the torus which is constructed by the addition of $f$ is
\begin{equation}
\label{eqn:eto}
	{\bf A}=-\Gamma \partial_{\bm\theta} V({\bf 0},{\bm\theta}),
\end{equation}
which is of order $\varepsilon$ since $ V({\bf 0},{\bm\theta})$ is of order $\varepsilon$.
There is a significantly large freedom in choosing the function $\Omega$. It is sufficient to have $\Omega( {\bf A},{\bm\theta})=1$ for $\Vert {\bf A}\Vert \leq \varepsilon$. For instance, $\Omega( {\bf A},{\bm\theta})=1$ would be a possible and simpler choice, however representing a long-range control since the control term $f({\bm\theta})$ would be applied on all phase space. On the opposite way, we can design a function $\Omega$ such that the control is localized around the created invariant torus. The support of the function is reduced to a small neighborhood of the torus ${\bf A}=-\Gamma \partial_{\bm\theta} V({\bf 0},{\bm\theta})$. The main advantage of this step function is that it needs fewer energy (only in the part of phase space where the control is localized) and also it does not change the other part of phase space, not altering the overall chaotic structure of the system. In particular, the dynamics is not changed in the region of the dominant resonances. Therefore a clever choice for the localization is
$$
\Omega=\Omega_{loc}\left(\Vert {\bf A}+\Gamma \partial_{\bm\theta} V({\bf 0},{\bm\theta})\Vert\right),
$$
where the parameters $\alpha=\sup \{x\in{\mathbb R}^+, \Omega_{loc}(x)= 1\}$ and $\beta=\sup \{x\in{\mathbb R}^+, \Omega_{loc}(x)\not= 0\}$ are small and $\Omega_{loc}$ is sufficiently smooth on ${\mathbb R}^+$ (for instance of class ${\mathcal C}^2$). One example of $\Omega_loc$ is depicted on Fig.~\ref{fig:omegaloc}.

{\em Remark~:} The amplitude of the control term is proportional to $\varepsilon^2$. However, if the aim of the control is to create a given invariant torus (e.g., specified by its frequency $\omega$), the control only makes sense if the invariant torus with frequency $\omega$ does not already exist in the original (uncontrolled) system. In other words, for $\varepsilon\leq\varepsilon_c(\omega)$ where $\varepsilon_c(\omega)$ is the threshold of break-up of the invariant torus, there is no need of control since the uncontrolled Hamiltonian has an invariant torus acting as a barrier (or an effective barrier) to diffusion. However, we point out that, below the critical threshold $\varepsilon_c(\omega)$, the specific location of the invariant torus in the uncontrolled system is not known exactly in general (this is in particular the case for the forced pendulum described below). With a small modification of the potential (adding the proposed control term), there is an exact formula~(\ref{eqn:eto}) giving the equation of the torus.  

In the next section, we will see on a specific example that the amplitude of the control term is small compared with the perturbation. We notice that it was the case for ${\bf E}\times {\bf B}$ drift motion which was considered in Refs.~\cite{guido1,guido2} and that it was also the case for the experimental control in the traveling wave tube (TWT)~\cite{twt}.

\section{Application to the forced pendulum}

\begin{figure}
\begin{center}
\epsfig{file=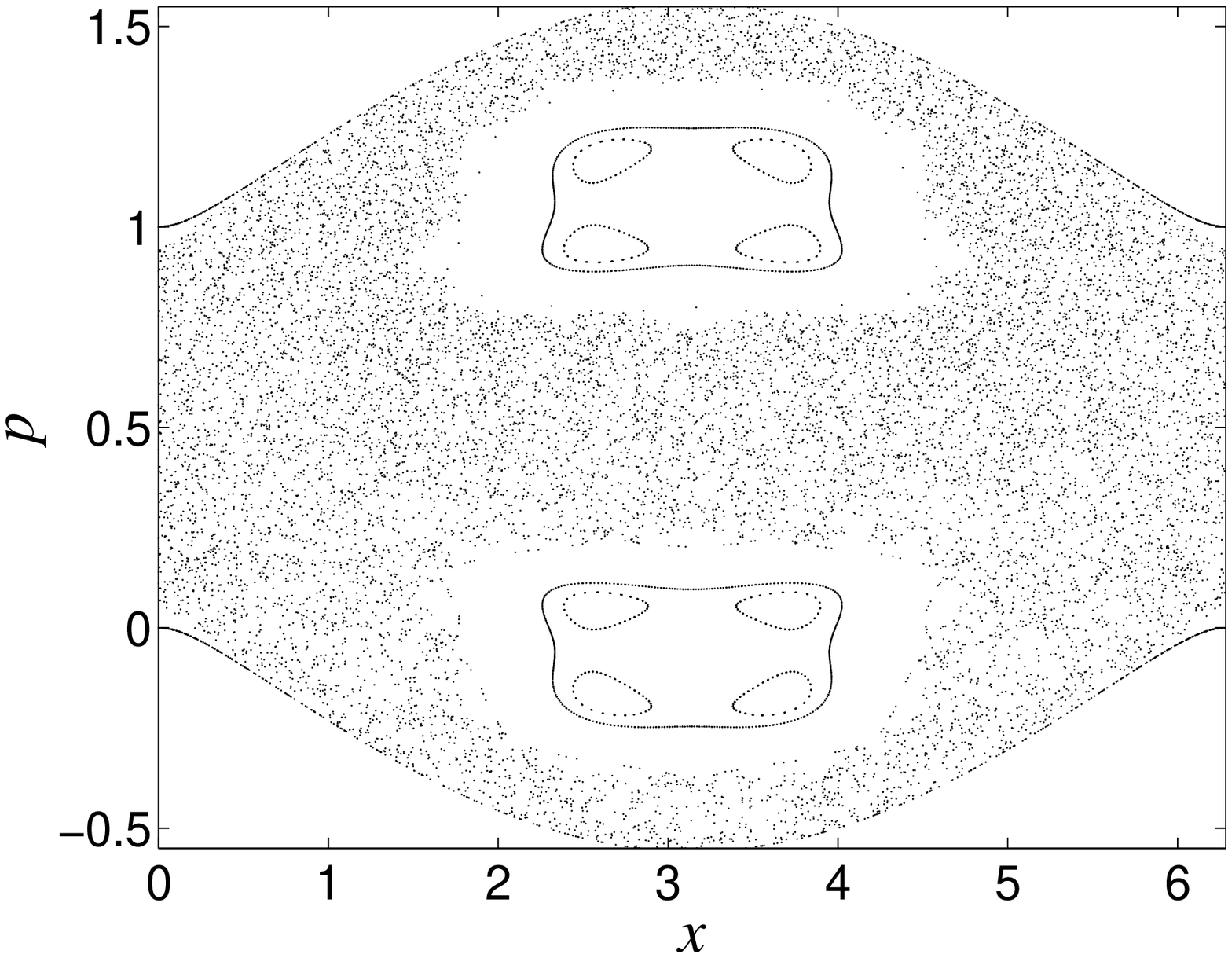,width=7.5cm,height=6.3cm}
\end{center}
\caption{Poincar\'e surface of section of Hamiltonian (\ref{eqn:fp}) with $\varepsilon=0.065$.}
\label{fig1}
\end{figure}

We consider the following forced pendulum model
\begin{equation}
\label{eqn:fp}
H(p,x,t)=\frac{1}{2}p^2+\varepsilon \left[ \cos x+\cos(x-t)\right].
\label{Hpend}
\end{equation}
If we consider the Chirikov's criterion, it leads to an estimate of the threshold of large scale chaos $\varepsilon=0.0625$~\cite{zav}.
In fact, the mechanism of the transition to chaos is much more complex than what it is hinted by this empirical criterion. Several methods have been developed since to get much deeper insight into the mechanism and also to get more accurate values of the transition to chaos (e.g., Greene's residue criterion~\cite{gree79,mack92}, frequency map analysis~\cite{lask92,lask99} and renormalization method~\cite{PRchan}). The large scale diffusion is due to the break-up of the last KAM torus. For this model, the last invariant torus is the one with frequency $\omega=(3-\sqrt{5})/2$ and its critical threshold is $\varepsilon\approx 0.02759$~\cite{PRchan}. A Poincar\'e surface of section is depicted on Fig.~\ref{fig1} for $\varepsilon=0.065$. As expected, a large chaotic zone takes place between the two primary resonances since all the KAM tori are broken for this value of $\varepsilon$.

\begin{figure}
\begin{center}
\epsfig{file=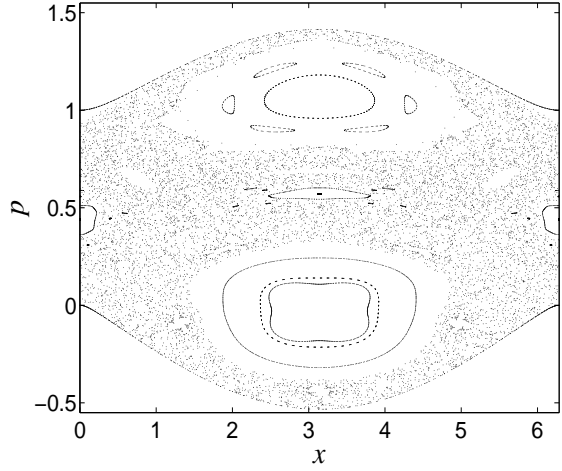,width=7.5cm,height=6.3cm}
\end{center}
\caption{Poincar\'e surface of section of Hamiltonian (\ref{eqn:Hcdumb}) with $\varepsilon=0.065$ and $\delta=0.5$.}
\label{fig4}
\end{figure}

A first and simple idea to control the system would be to reduce the amplitude of one of the two primary resonances by considering the following control term which is of order $\varepsilon$~:
$$
f_d(x,t)=-\delta \varepsilon \cos (x-t),
$$
where $\delta$ is an additional control parameter.
In this case the control term becomes
\begin{equation}
\label{eqn:Hcdumb}
H_c(p,x,t)=\frac{1}{2}p^2+\varepsilon \left[ \cos x+(1-\delta)\cos(x-t)\right].
\end{equation}
Figure~\ref{fig4} shows a Poincar\'e section of Hamiltonian~(\ref{eqn:Hcdumb}) for $\varepsilon=0.065$ and $\delta=0.5$. As expected the size of the upper resonance decreases but this significant reduction of a primary resonance (which results in a decrease of the Chirikov parameter) does not succeed in creating an invariant torus in between the two resonances. 

In order to create an invariant torus, one needs to choose $0.7\leq \delta\leq 1$ which requires too much energy for the control. In addition, even if $\delta$ satisfies the required condition, there is no explicit formula for the equation of the created invariant torus which is crucial in order to localize the control term and hence to reduce drastically the amount of energy necessary for the control. 
Therefore one has to design another strategy which has two main goals~:
\begin{itemize}
\item find a small control term $f$ which creates an invariant torus between the two resonances,
\item know explicitly the location of the created invariant torus in order to localize the control.
\end{itemize} 
In what follows, we show that the method proposed in Ref.~\cite{vitt04} gives a much smaller control term (of order $\varepsilon^2$) such that it creates an invariant torus whose equation is explicitly known. 

In order to compute this apt control term, this Hamiltonian with 1.5 degrees of freedom is mapped in the usual way into an autonomous Hamiltonian with two degrees of freedom by considering that $t \mbox{ mod }2\pi$ is an additional angle variable. We denote $E$ its conjugate action.
The autonomous Hamiltonian is
\begin{equation}
\label{eqn:H2dof}
H=E+\frac{p^2}{2}+\varepsilon \left[\cos x+\cos(x-t)\right].
\end{equation}
The aim of the localized control is to modify locally Hamiltonian~(\ref{eqn:H2dof}) and therefore (\ref{eqn:fp}), in order to reconstruct an invariant torus with frequency $\omega$. We assume that $\omega$ is sufficiently irrational (diophantine) in order to fulfill the hypotheses of the KAM theorem. First, the momentum $p$ is shifted by $\omega$ in order to define a localized control in the region $p\approx 0$, i.e.\ to get the invariant torus located near $p\approx\omega$ for Hamiltonian~(\ref{eqn:H2dof}) for $\varepsilon$ sufficiently small. The operator $\Gamma$ is defined from the integrable part of the Hamiltonian which is linear in the actions $(E,p)$~:
$$H_0(E,p)=E+\omega p,$$
and Hamiltonian~(\ref{eqn:H2dof}) is
$$
H=H_0+V,
$$ 
where 
\begin{equation}
\label{eqn:Vfp}
V(p,x,t)=\varepsilon \left[\cos x+\cos(x-t)\right]+\frac{p^2}{2}.
\end{equation}
The action of $\Gamma$ on a function $U$ of $p$, $x$ and $t$ defined as
$$
U(p,x,t)=\sum_{(k_1,k_2) \in {\mathbb Z}^2} U_{k_1,k_2}(p) {\mathrm e}^{i(k_1 x+k_2 t)},
$$
is given by
$$ 
\Gamma U =\sum_{(k_1,k_2)\not= (0,0)} \frac{U_{k_1,k_2}}{i(\omega k_1+k_2)} {\mathrm e}^{i(k_1 x+k_2 t)}.
$$
For the forced pendulum~(\ref{eqn:fp}), we have
$$
\Gamma \partial_{x}V(0,x,t)=\varepsilon \left[\frac{\cos x}{\omega}+\frac{\cos(x-t)}{\omega-1}\right].
$$
The control term given by Eq.~(\ref{eqn:exf}) is equal to
\begin{equation}
	f(x,t)=-\frac{1}{2}(\Gamma \partial_{x} V(0,x,t)) ^2=-\frac{\varepsilon^2}{2}\left( \frac{\cos x}{\omega}+\frac{\cos(x-t)}{\omega-1}\right)^2.
	\label{eqn:fpa}
\end{equation}
With the addition of the control term given by Eq.~(\ref{eqn:fpa}), the controlled Hamiltonian has an invariant torus whose equation is
\begin{equation}
\label{eqn:torefp}
p_0(x,t)=\omega-\varepsilon \left[\frac{\cos x}{\omega}+\frac{\cos(x-t)}{\omega-1}\right].
\end{equation}
The controlled Hamiltonian is given by
\begin{equation}
\label{eqn:Hc}
H_c(p,x,t)=\frac{1}{2}p^2+\varepsilon \left[ \cos x+\cos(x-t)\right]-\frac{\varepsilon^2}{2}\left( \frac{\cos x}{\omega}+\frac{\cos(x-t)}{\omega-1}\right)^2 \Omega(p,x,t),
\end{equation}
where we will consider two cases for $\Omega$:
\begin{eqnarray*}
	&& \Omega(p,x,t)=1,\\
	&& \Omega(p,x,t)=\Omega_{loc}(\vert p-p_0(x,t)\vert),
\end{eqnarray*}
and $\Omega_{loc}:{\mathbb R}^+\to [0,1]$ will be specified below.
We notice that we can subtract to the control term $f$ any function which is constant in $t$. In particular, we can substract to the control term its Fourier components with ${\bf k}=(0,k)$ for any $k\in {\mathbb Z}$. This can be seen by looking at Hamilton's equations and, since $\Omega$ is constant (equal to 1) in the region near the invariant torus, the equations of motion are the same in this region with or without the function depending only on time. Therefore the control term can be reduced to three Fourier modes
$$
f(x,t)=-\varepsilon^2\left[ \frac{\cos 2x}{4\omega^2}+\frac{\cos 2(x-t)}{4(1-\omega)^2}+\frac{\cos (2x-t)}{2\omega(\omega-1)} \right].
$$ 

{\em Remark~:} We consider a forced pendulum when there are two parameters, one for the size of the primary resonances and another for the distance between them~:
\begin{equation}
\label{eqn:fpd}
H(p,x,t)=\frac{p^2}{2}+\varepsilon_d [ \cos x +\cos (x-dt)].
\end{equation}
By rescaling the actions and by rescaling time, this Hamiltonian reduces to the forced pendulum~(\ref{eqn:fp}) with $\varepsilon=\varepsilon_d/d^2$. If we fix the value of $\varepsilon$ to the value of the break-up of the last KAM surface $\varepsilon_c\approx 0.02759$ (or to another value greater than this one, representing the degree of chaoticity in the model), then we have $\varepsilon_d \sim d^2 $. If we compute the control term in the original variables for a rescaled frequency $d\omega$ (since we want to keep $\omega\in ]0,1 [$), it leads to the following control term~:
$$
f_d(x,t)=-\frac{\varepsilon_d^2}{d^2}\left( \frac{\cos x}{\omega}+\frac{\cos(x-t)}{\omega-1}\right)^2.
$$ 
If we fix the degree of chaoticity $\varepsilon$ constant and eliminate the parameter $d$, the amplitude of the control term is proportional to $\varepsilon \varepsilon_d$. Therefore the control term is of order $\varepsilon_d$ for the model~(\ref{eqn:fpd}). The fact that $f_d$ is also proportional to $\varepsilon$ which is in general small, makes the control term for Hamiltonian~(\ref{eqn:fpd}) very small compared with the size of the perturbation.\\

We consider the following norm of a function $U(x,t)$ as
$$
\Vert U\Vert=\max_{(x,t)\in{\mathbb T}^2}\vert U(x,t)\vert.
$$
The norm of the perturbation is 
$$
\Vert V\Vert = 2\varepsilon,
$$
and the norm of the control term given by Eq.~(\ref{eqn:fpa}) is
$$
\Vert f\Vert = \frac{\varepsilon^2}{4\omega^2(1-\omega)^2}.
$$
Therefore the ratio $r$ between the norm of the control term and the perturbation is given by
$$
r=\frac{\varepsilon}{8\omega^2(1-\omega)^2}.
$$ 
The ratio $r$ is smaller when $\omega$ is close to 1/2 and large when it is close to the primary resonances $\omega\approx 0$ or 1. For instance, for $\varepsilon=0.034$ and $\omega$ close to $1/2$, this ratio is approximately 7\%. 
For the numerical computations we have chosen $\omega=(3-\sqrt{5})/2$ and $\varepsilon=0.065$. For these values of the parameters, the ratio $r$ is about 15\%. 

\begin{figure}
\begin{center}
\epsfig{file=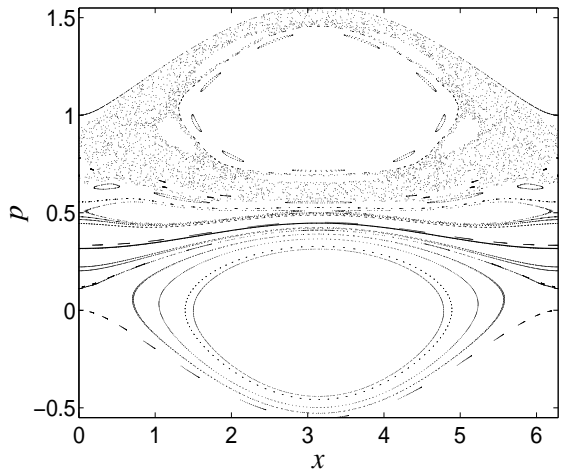,width=7.5cm,height=6.3cm}
\end{center}
\caption{Poincar\'e surface of section of the controlled Hamiltonian (\ref{eqn:Hc}) with $\varepsilon=0.065$ and $\Omega=1$.}
\label{figCG}
\end{figure}

A Poincar\'e section of the controlled Hamiltonian~(\ref{eqn:Hc}) for $\varepsilon=0.065$ shows that a lot of invariant tori are created with the addition of the control term precisely in the lower region of phase space where the localization has been done (see Fig.~\ref{figCG}).\\
The next step is to localize $f$ given by Eq.~(\ref{eqn:fpa}) around the invariant torus created by $f$ whose equation is given by Eq.~(\ref{eqn:torefp}). More precisely, we have chosen $\Omega_{loc}(x)=1$ for $x \leq \alpha$, $\Omega_{loc}(x)=0$ for $x \geq \beta$ and a third order polynomial for $x \in ]\alpha,\beta[$ for which $\Omega_{loc}$ is a $C^1$-function, i.e.\ $\Omega_{loc}(x)=1-(x-\alpha)^2(3\beta-\alpha-2x)/(\beta-\alpha)^3$. We have chosen $\alpha= 10^{-2}$ and $\beta=1.5\alpha$. The support in momentum $p$ of the localized control is of order $10^{-2}$ compared with the support of the global control which is of order 1. Therefore the localized control term requires less than 0.5\% of the energy of the perturbation.

\begin{figure}
\begin{center}
\unitlength 1cm
\begin{picture}(15,6.3)(0,0)
\put(0,0){\epsfig{file=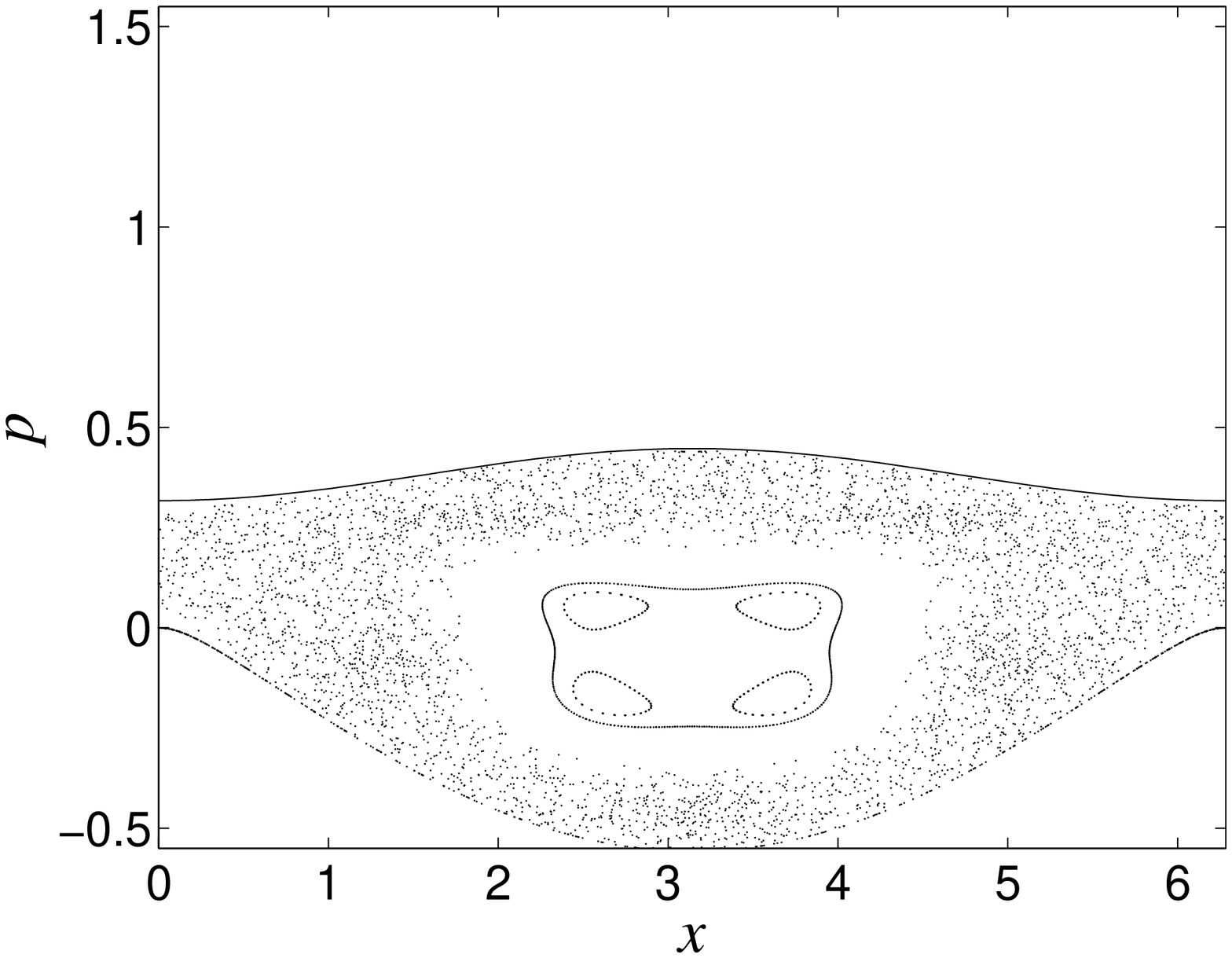,width=7.5cm,height=6.3cm}}
\put(8,0){\epsfig{file=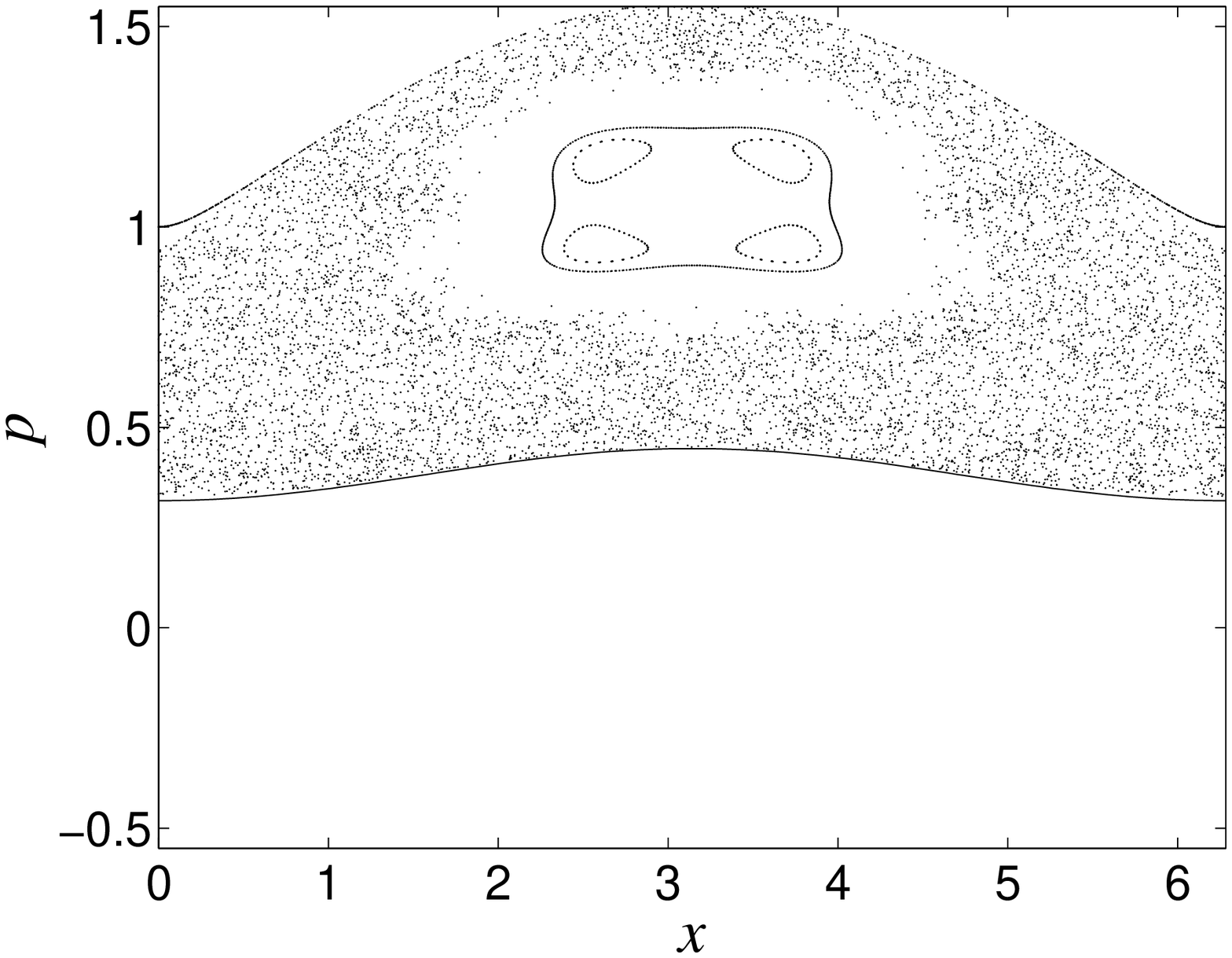,width=7.5cm,height=6.3cm}}
\end{picture}
\end{center}
\caption{Poincar\'e surfaces of section of the controlled Hamiltonian (\ref{eqn:Hc}) with $\varepsilon=0.065$ and initial conditions launched from $(a)$ below the invariant torus and $(b)$ from above for $\Omega=\Omega_{loc}$ as defined in the paper.}
\label{fig2}
\end{figure}

Figure~\ref{fig2} shows that the phase space of the controlled Hamiltonian is very similar to the one of the uncontrolled Hamiltonian. We notice that there is in addition an isolated invariant torus. A complete barrier to diffusion is then created.
Since the control is robust, we notice that there is also the possibility of reducing the amplitude of the control (by a few percent, see Ref.~\cite{guido3}) and still get an invariant torus of the desired frequency for a perturbation parameter $\varepsilon$ significantly greater than the critical value in the absence of control. Also, there is the possibility in selecting appropriate Fourier coefficients of the control term and still get the selected invariant torus.

\begin{figure}
\begin{center}
\unitlength 1cm
\begin{picture}(15,6.3)(0,0)
\put(0,0){\epsfig{file=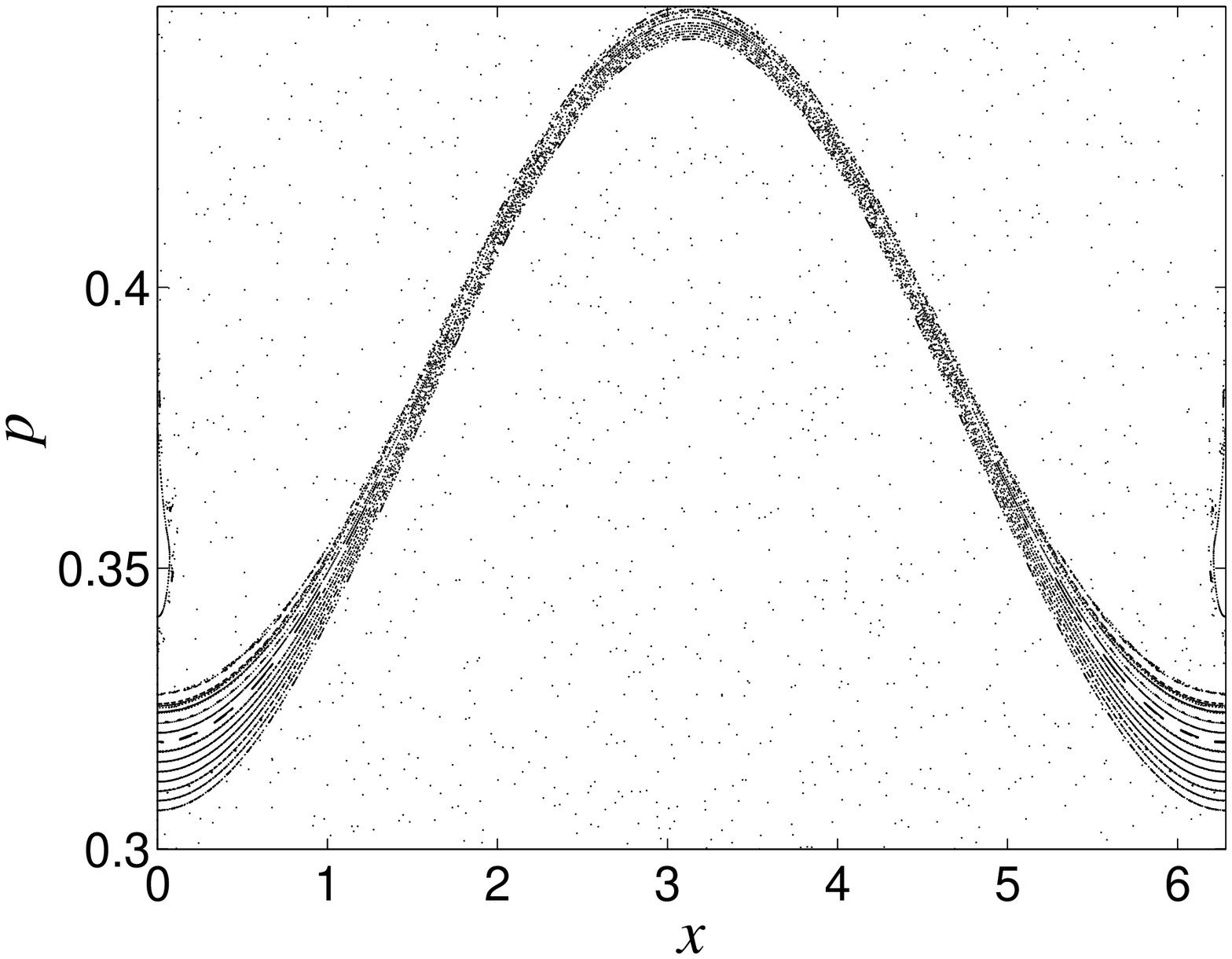,width=7.5cm,height=6.3cm}}
\put(8,0){\epsfig{file=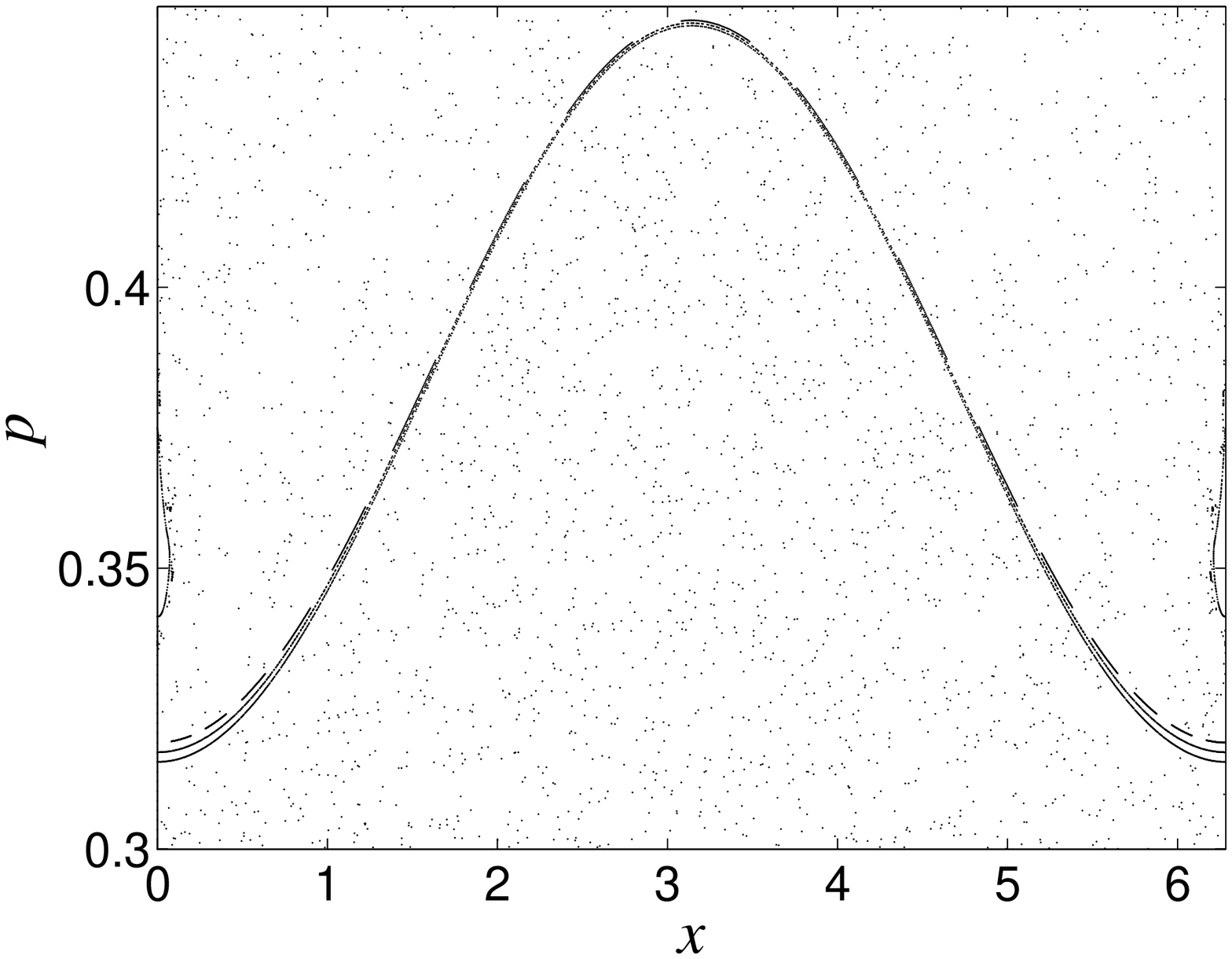,width=7.5cm,height=6.3cm}}
\end{picture}
\end{center}
\caption{Poincar\'e surface of section of the controlled Hamiltonian (\ref{eqn:Hc}) with $\varepsilon=0.065$ and $\Omega=\Omega_{loc}$ as defined in the paper~: local dynamics around the recreated invariant torus for $\alpha=0.01$ (left panel) and for $\alpha=0.002$ (right panel).}
\label{figCGloc}
\end{figure}

Figure~\ref{figCGloc} shows a Poincar\'e surface of section of the controlled Hamiltonian~(\ref{eqn:Hc}) in a small region around the invariant torus given by Eq.~(\ref{eqn:torefp}). In addition to the creation of the invariant torus, we notice that the effect of the control term is also to regularize a small neighborhood of the invariant torus. We have represented this local dynamics for two different values of the parameter $\alpha$~: for $\alpha=0.01$ on the left panel and for $\alpha=0.002$ on the right panel. We notice that this regularized neighborhood can be chosen arbitrarily small (by choosing arbitrarily small values for $\alpha$). Moreover, we notice that this barrier persists to arbitrarily large values of the coupling parameter $\varepsilon$.

\section*{Acknowledgments}
We acknowledge useful discussions with J R Cary, F Doveil, Ph Ghendrih, J Laskar, X Leoncini, A Macor, M Pettini and the Nonlinear Dynamics group at CPT.
This work is supported by Euratom/CEA 
(contract EUR 344-88-1 FUA F).

\end{document}